\documentclass[rough]{ametsocjmk}

\usepackage{graphicx}
\usepackage{amsmath}
\usepackage{pslatex}
\usepackage{multirow}

\renewcommand{\vec}[1]{\mbox{${\mathbf #1}$}}

\journalid{}

\authoraddr{Sushil Shetty, 6116 Etcheverry Hall, University of California, 
Berkeley, CA 94720\\E-mail: sushil@newton.berkeley.edu}
 
\begin{document}

\title{On the Interaction of Jupiter's Great Red Spot and Zonal Jet Streams}

\author[1]{Sushil Shetty}
\author[2]{Xylar S. Asay-Davis}
\author[1,2]{Philip S. Marcus}

\affil[1]{Dept. of Mechanical Engineering, University of California, 
Berkeley, CA 94720}
\affil[2]{Applied Science and Technology Program, University of California, 
Berkeley, CA 94720}

\begin{abstract}
\noindent In this paper, Jupiter's Great Red Spot (GRS) is 
used to determine properties of the Jovian atmosphere that cannot 
otherwise be found.  
These properties include the potential vorticity of the GRS and its 
neighboring jet streams, the shear imposed on the GRS by the jet streams, 
and the vertical entropy gradient (i.e., Rossby deformation radius).  
The cloud cover of the GRS, which is often used to define the 
GRS's area and aspect ratio, is found to differ significantly from the 
region of the GRS's potential vorticity anomaly.
The westward-going jet stream to the north of the GRS and the eastward-going 
jet stream to its south are each found to have a large potential vorticity ``jump''.  
The jumps have opposite sign and as a consequence of their 
interaction with the GRS, the shear imposed on the GRS is reduced.  
The east-west to north-south aspect ratio of the GRS's potential vorticity 
anomaly depends on the ratio of the imposed shear to the strength of the 
anomaly.  
The aspect ratio is found to be $\approx$2:1, 
but without the opposing jumps it would be much greater.  
The GRS's high-speed collar and quiescent interior require that the 
potential vorticity in the interior be approximately half that in 
the collar.  No other persistent geophysical vortex has a significant 
minimum of potential vorticity in its interior and laboratory 
vortices with such a minimum are unstable.  
\end{abstract}

\section{Introduction}
\label{sec:intro}

Only a relatively thin ($\sim$10~km) outer layer of Jupiter's atmosphere 
containing the visible clouds and vortices is accessible by direct 
observation.  Most of the details of the underlying layers, such as the 
vertical stratification, must therefore be determined indirectly.  
In this paper, we present one such indirect method.  In particular, 
we use the observed velocity field of a persistent Jovian vortex to 
determine quantities relevant to both outer and underlying layers.  
These quantities are the potential vorticity of the vortex, the 
potential vorticity of the neighboring jet streams, the flow in the 
underlying layers, and the Rossby deformation radius $L_r$, 
which is a measure of the vertical stratification.  
We demonstrate the method using {\it Voyager}~1 observations 
of the Great Red Spot (GRS).  

We are not the first to use the GRS velocity field as a probe of the 
Jovian atmosphere \citep{dowling88, dowling89, cho01}.  
However, our approach differs from previous ones in several significant 
respects.  First, the GRS velocity field is sufficiently noisy that 
we do not, unlike in previous analyses, take spatial derivatives of the 
velocity to compute potential vorticity.  
Instead, we solve the {\it inverse problem}:
We identify several ``traits'' of the GRS velocity field, where a 
trait is a feature of the velocity field that is unambiguously 
quantifiable from the noisy data.  
We then construct a model for the flow and determine ``best-fit'' values 
for the model parameters such that the model velocity field 
reproduces the observed traits.  
Furthermore, for a given set of parameter values, we construct the model 
velocity field so that it is an {\it exact steady solution} of the 
equations that govern the flow.  
For the {\it Voyager}~1 data, we find that a best-fit model 
(i.e., a trait-reproducing steady solution) determined in this manner 
agrees with the entire GRS velocity field to within the 
observational uncertainties.  
 
A second way in which our study differs from previous ones is that we 
explicitly compute the interaction between the GRS and its neighboring 
jet streams.  We show that the interaction controls the aspect ratio of the 
GRS's potential vorticity anomaly, 
which is relevant to recent observations that show the aspect ratio of the 
GRS's cloud cover to be a function of time \citep{simon02}.  
The changing cloud cover, if symptomatic of changes in the GRS's 
potential vorticity anomaly, would be indicative of a change in the 
interaction and a corresponding change in the best-fit values of the 
parameters that govern the interaction.  
Finally, in this study, we quantify the relationship between individual 
traits and individual parameters.  When a trait is nearly 
independent of all parameters except for one or two, a clear physical 
understanding is obtained between ``cause'' (a model parameter) 
and ''effect'' (a GRS trait). 

Our philosophy is to use a model with the fewest free parameters that is 
an exact steady solution to the least complex governing equation, yet can
still reproduce the observed velocity to within its uncertainties. 
The danger of more complex models is that they have larger degrees of freedom. 
By varying parameters they can fit the observed velocity but 
misidentify the relevant physics.  
For the {\it Voyager}~1 data considered here, we use the 
1.5--layer reduced gravity quasigeostrophic (QG) equations and a model 
with nine free parameters.  
The {\it Voyager}~1 data can be reproduced with this model and 
{\it does not warrant} models with more free parameters or 
governing equations with more complexity.  

The rest of the paper is organized as follows.  In $\S$2 we determine 
the GRS velocity field from {\it Voyager}~1 observations and then identify 
traits of the velocity.  In $\S$3 we review the governing equations and 
describe a decomposition of the flow around the GRS into a near-field, 
a far-field, and an interaction-field.  
In $\S$4 we define the model and list its free parameters.  
In $\S$5 we determine best-fit parameter values, i.e., parameter values 
for which the model reproduces the traits.  
In $\S$6 we discuss the physical implications of the best-fit model, 
and in $\S$7 conclude with an outline for future work.  

\section{GRS velocity field}\label{sec:observations}

\subsection{Determination of GRS velocity}

\begin{figure}
\begin{center}
\includegraphics[width=16.5cm]{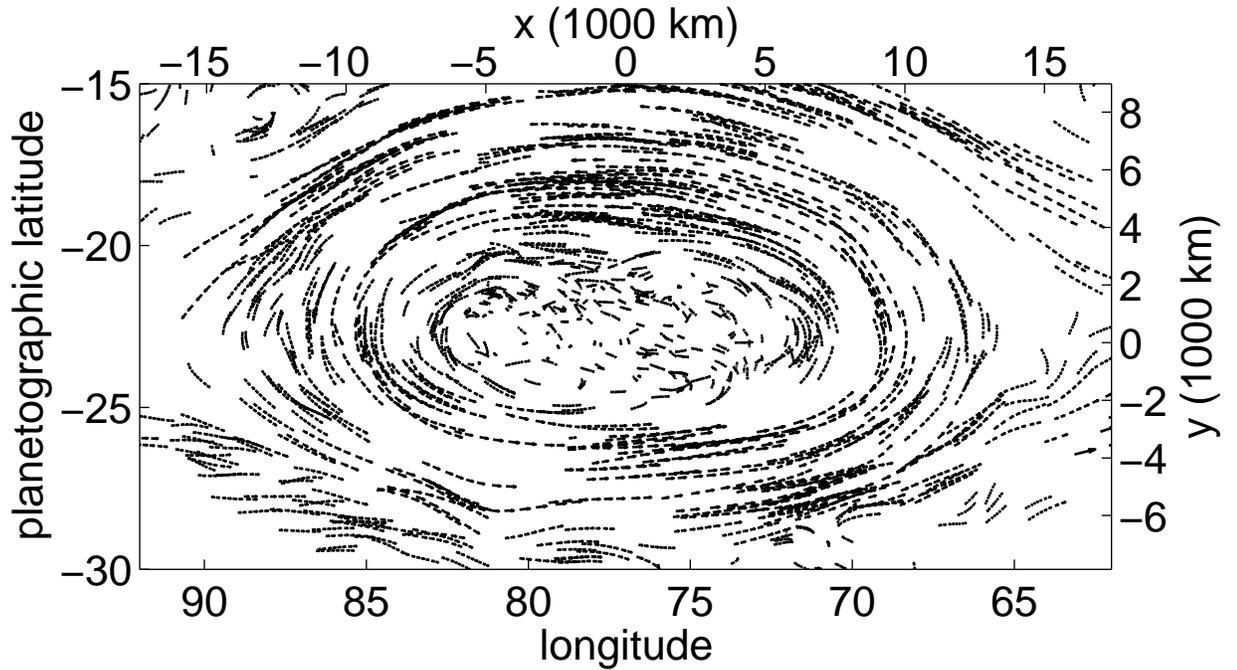}
\caption{Velocity vectors of the GRS with respect to System III as determined 
from {\it Voyager}~1 images.  
The velocities were determined by dividing the displacement of a cloud feature 
in a pair of images by the time between the two images \citep{mitchell81}, 
and then correcting for the fact that cloud trajectories over typical 
image separation times are not straight lines (see $\S$2a).  
The $23^\circ$S latitude is defined to be the principal east-west 
(E--W) axis, and the $77^\circ$W longitude is defined to be the principal 
north-south (N--S) axis.  
\label{fig:vectors}}
\end{center}
\end{figure}

In \citet{mitchell81}, {\it Voyager}~1 images were used to determine 
the GRS velocity field by dividing the displacement of a cloud feature in a 
pair of images by the time interval between the images 
(typically one Jovian day or $\approx$10 hours).  
The cloud features were identified by hand rather than by 
an automated approach such as Correlation-Image-Velocimetry (CIV: 
Fincham and Spedding 1997), and may therefore contain spurious velocities on 
account of misidentifications.  
Furthermore, dividing cloud displacement by time does not account for the 
curvature of a cloud trajectory, since in ten hours, a cloud feature in the 
high-speed collar travels almost a third of the way across the GRS.  
However, due to the unavailability of the original navigated images, 
we use the Mitchell velocities, but remove some of the errors 
by a procedure described in appendix~A.  
The procedure leads to the removal of 220 of the original 1100 measured 
cloud displacements and the addition of 7100 synthetic measurements.  
The net result is that the uncertainty in the velocity field is reduced 
from $\approx9$~m~s$^{-1}$ to $\approx7$~m~s$^{-1}$.  
Fig.~\ref{fig:vectors} shows the processed GRS velocity field.  
Consistent with previous analyses, the velocity field shows a 
quiescent core and high-speed collar.  
The inner part of the collar has anticyclonic vorticity, and the 
outer part has cyclonic vorticity.
The peak velocities in the collar are $140\pm 7$~m~s$^{-1}$ and the peak 
velocities in the core are $7 \pm 7$~m~s$^{-1}$.  

\begin{figure}
\begin{center}
\includegraphics[width=16.5cm]{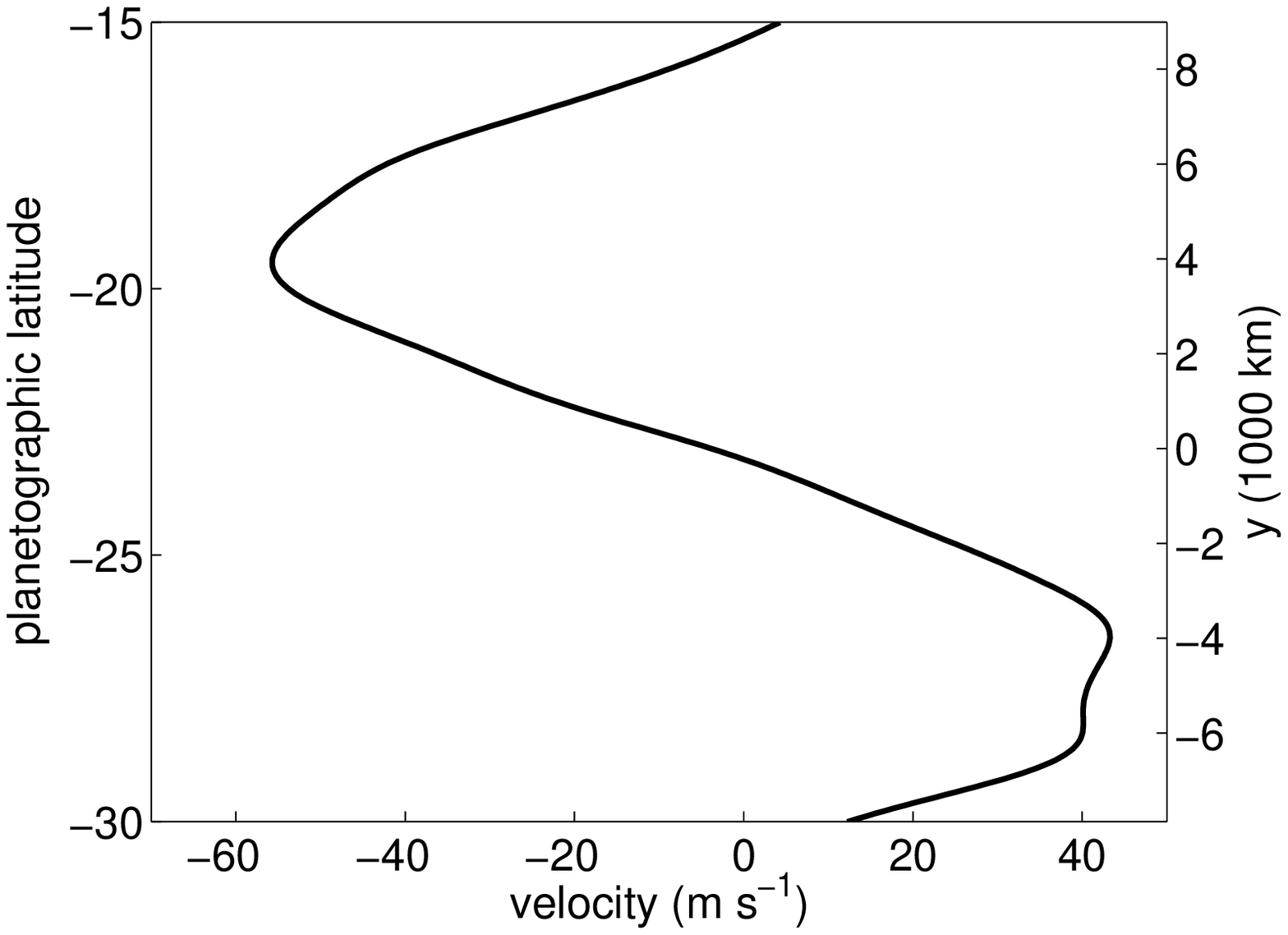}
\caption{Averaged zonal velocity ${\bf v}^{\infty}$ in System III 
\citep{limaye86}.  
\label{fig:limaye}}
\end{center}
\end{figure}

The GRS is embedded in a zonal (east-west) flow.  
The zonal mean of this flow, averaged over 142 Jovian days, 
was computed from {\it Voyager}~2 images \citep{limaye86}, and 
is shown in Fig.~\ref{fig:limaye}.  
Between $15^\circ$S\footnote[1]{In this paper all latitudes 
are planetographic.} and $30^\circ$S, the profile is characterized by a 
westward-going jet stream that peaks at $\approx19.5^\circ$S, 
and an eastward-going jet stream that peaks at $\approx26.5^\circ$S.  
The uncertainty in the profile is 7~m~s$^{-1}$.  
(Note that most likely due to navigational errors \citep{limaye86}, 
the published profile must be shifted north by $0.5^{\circ}$ so as to be 
consistent with the navigated latitudes of {\it Voyager}~1 in 
Fig.~\ref{fig:vectors}.)
The GRS was observed to drift westward at a rate of 3--4~m~s$^{-1}$ with 
respect to System III during the {\it Voyager} epoch \citep{dowling88}.   

\subsection{Pitfalls to be avoided when analyzing GRS velocity}

We do not compute quantities by taking spatial derivatives of 
the velocity data, as this tends to amplify small length 
scale noise.  For example, in the high-speed collar, we found that 
the uncertainty in vorticity obtained by differentiating the velocity 
is $\approx35\%$ of the maximum vorticity.  
If vorticities must be found, it is usually better to integrate the 
velocity to obtain a circulation and then divide by an area to obtain a 
local average vorticity.  
We also do not average the velocity locally, which is a standard way of 
reducing noise.  For example, if the GRS velocity is averaged over 
length scales greater than $2^\circ$, the peak velocities and vorticities 
are severely diminished.  This is due to the fact that an 
averaging length of $2^\circ$ is too large; it corresponds to 
$\approx2500$~km, which is the length scale over which the velocity 
changes by order unity (cf. the width of the high-speed collar).  
Finally, we do not obtain a quantity by adding two numbers of similar 
magnitude but opposite sign, so that the resulting sum is of order or smaller 
than the uncertainty in each of the numbers being summed.
For example, if the velocity is assumed to be divergence-free, the vertical 
derivative of the vertical velocity $\partial v_z/\partial z$ can be obtained 
by computing the negative of the horizontal divergence 
$\partial v_x/\partial x + \partial v_y/\partial y$ .  
However, a simple scaling argument shows that the horizontal divergence 
is smaller than each partial derivative term separately, and in particular, 
is of the same order as the uncertainty in each term 
(which is relatively large because the terms are derivatives of noisy data).  
Thus $\partial v_z/\partial z$ computed in this fashion would have order 
unity uncertainties.

\subsection{Traits of GRS velocity}

\begin{figure}
\begin{center}
\includegraphics[width=16.5cm]{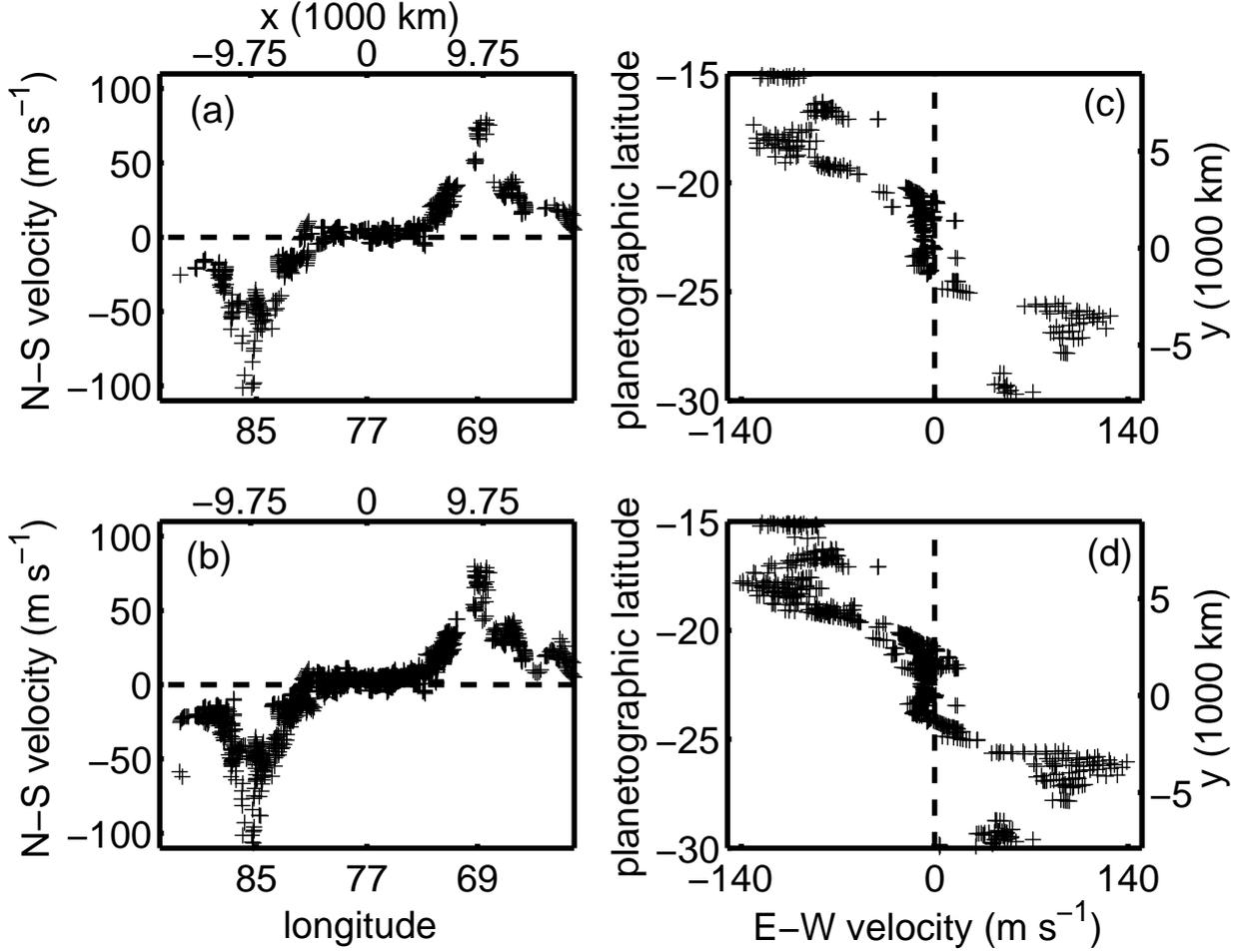}
\caption{Parts (a) and (b) show the north-south component of the 
velocity from Fig.~\ref{fig:vectors}, for points that lie within 
$0.7^{\circ}$ and $1.4^{\circ}$ respectively, of the principal east-west axis.  
Parts (c) and (d) show the east-west component of the velocity 
for points that lie within $0.7^{\circ}$ and $1.4^{\circ}$ respectively, 
of the principal north-south axis.  
Parts (a), (b), (c), and (d) contain 903, 1992, 586, and 1163 points 
respectively.  
\label{fig:axesTiePoints}}
\end{center}
\end{figure}

The traits that we consider are derived from the north-south (N--S) 
velocity along the principal east-west (E--W) axis and from the 
east-west (E--W) velocity along the principal north-south (N--S) axis.  
The E--W and N--S principal axes are defined to be the $23^\circ$S latitude 
and the 77$^\circ$W longitude respectively.  
The point of intersection of the principal axes is roughly the 
centroid of the GRS as inferred from its clouds.  
The velocity profiles along the axes are shown in Fig.~\ref{fig:axesTiePoints}.
To better understand the pitfalls of local averaging, 
Figs.~\ref{fig:axesTiePoints}a and 3c show the velocities from Fig.~1 
for points that lie within $\pm 0.7^{\circ}$ 
of the axes, while Figs.~3b and 3d show the velocities that lie within 
$\pm 1.4^{\circ}$ of the axes.  
The axes labels $x$ and $y$ in the figure denote local E--W and N--S 
cartesian coordinates.  
Based on the figure, we define the following to be {\it traits} of the 
velocity field:  
(1) the northward-going jet and southward-going jet in 
Figs.~\ref{fig:axesTiePoints}(a)--(b) that peak at $x=\pm9750\pm500$~km 
respectively and have peak magnitude $V^{NS}_{\rm max} = 95\pm 7$~m~s$^{-1}$, 
(2) the small magnitude N--S velocity in $|x|\leq6000$~km, 
(3) the eastward-going jet and westward-going jet in 
Figs.~\ref{fig:axesTiePoints}(c)--(d) that peak at 
$y=-3500\pm500$~km and $y=5500\pm500$~km respectively, 
and have peak magnitude $140\pm7$~m~s$^{-1}$, 
(4) the small magnitude E--W velocity in $|y|\leq2000$~km.  
Traits (2) and (4) illustrate the quiescent interior of the GRS.  
Traits (1) and (3) illustrate the high-speed collar.   
The uncertainties in peak velocities are from the global estimate in $\S2$a.  
The uncertainties in peak locations are not rigorous.  
They are from an estimate of the spatial scatter of points near 
the peak location.  
Henceforth, traits (1) and (2) will be referred to as the N--S velocity 
traits, and traits (3) and (4) as the E--W velocity traits.  

\section{Governing equations}
\label{sec:govEqns}

\subsection{1.5--layer reduced gravity QG approximation}

We do not model the whole sphere, but only a domain that extends from 
$15^\circ$S to $30^\circ$S.  For the flow in this domain, we adopt the 
1.5--layer reduced gravity QG equations on a beta-plane \citep{ingersoll81}.  
A derivation of the equations and the justification for their use 
can be found in \citet{dowling95}.  
Briefly, the layers correspond to an upper layer (also called 
``weather'' layer) of constant density $\rho_1$ and a much deeper lower 
layer of constant density $\rho_2 > \rho_1$.  
The upper layer contains the visible clouds and vortices while the lower layer 
contains a steady zonal flow. 
The two layers are dynamically equivalent to a single layer with rigid 
bottom topography $h_b$ and effective gravity 
$g\equiv g_J(\rho_2-\rho_1)/\rho_2$, where $g_J$ is the 
true gravity in the weather layer, and the bottom topography is a 
parametrization of the flow in the lower layer.  
The governing equation for the system advectively conserves a potential 
vorticity $q$:
\begin{equation}
\frac{Dq}{Dt}\equiv\left(\frac{\partial}{\partial t}+
\vec{v}\cdot \nabla\right)q=0,
\label{eqn:qg}
\end{equation}
\begin{equation}
q(x,y,t)\equiv\nabla^2\psi-\frac{\psi}{L_r^2}+\frac{gh_{b}(y)}{L_r^2f_0} + 
\beta y.
\label{eqn:q}
\end{equation}
Here $x$ and $y$ are the local E--W and N--S coordinates, 
$\psi$ is the streamfunction, 
$\vec{v}\equiv{\bf\hat{z}}\times{\bf\nabla}\psi$ is the weather 
layer velocity,
${\bf\hat{z}}$ is the local vertical unit vector,  
$\beta$ is the local gradient of the Coriolis parameter $f(y)$, 
$f_0$ is the local value of $f(y)$, and $L_r$ is the local Rossby 
deformation radius.  
Since $g$ appears only in combination with $h_b$, we shall refer to $gh_b$ 
as the bottom topography.  
Restricting $gh_b$ to be a function of $y$ alone restricts the flow 
in the lower layer to be steady and zonal with no vortices.  
The case $gh_b=0$, or ``flat'' bottom topography, corresponds to the lower 
layer being at rest in the rotating frame.   

\subsection{The near-field}

We assume that the GRS is a compact region (or patch) of anomalous potential 
vorticity.  
We denote the potential vorticity distribution of the GRS by 
$q^{GRS}(x,y)$, and for reasons that will become clear below, we refer to 
$q^{GRS}$ as the near-field.  
We define the streamfunction and velocity of the near-field to be:
\begin{equation}
q^{GRS}(x,y) \equiv \left(\nabla^2-1/L_r^2\right)\psi^{GRS}(x,y)
\label{eqn:justgrs}
\end{equation}
\begin{equation}
\vec{v}^{GRS} \equiv {\bf\hat{z}}\times{\bf\nabla}\psi^{GRS}. 
\label{eqn:vjustgrs}
\end{equation}
The velocity induced by a QG patch decays as exp($-r/L_r$), 
where $r$ is the distance from the patch boundary \citep{marcus90, marcus93}.  
Due to the exponential decay of velocity, a region of fluid 
that contains the patch and whose average radius is a few $L_r$ greater 
than the patch radius will have a circulation (or integrated vorticity) 
that is approximately zero.  
It would therefore be incorrect under the QG approximation to refer to 
the vorticity of the GRS as anticyclonic since its net vorticity is zero.  
On the other hand, the {\it potential vorticity} of the GRS is 
anticyclonic, as is the vorticity of most of its quiescent interior 
and the inner portion of its high-speed collar, but the vorticity of 
the outer portion of its collar is cyclonic 
(which is easily verified by noting that the azimuthal velocity in that 
region falls off faster than $1/r$). 

\subsection{The far-field}

The region of flow two or three deformation radii distant from the 
patch boundary, where the influence of the GRS is small, is 
defined to be the far-field.  
We assume the far-field flow to be zonal and independent of time and 
longitude.  
Eq.~\ref{eqn:q} then provides a relationship between the 
far-field velocity $\vec{v^\infty}\equiv v_x^{\infty}(y){\bf\hat{x}}$, 
and the far-field potential vorticity $q^{\infty}(y)$:
\begin{equation}
q^{\infty}(y) = 
\left(\frac{d^2}{dy^2}-\frac{1}{L_r^2}\right)\psi^{\infty}(y)+
\frac{gh_{b}(y)}{f_0L_r^2}+\beta y.
\label{eqn:qInf}
\end{equation}
For all calculations in this paper, $v_x^{\infty}(y)$ is prescribed 
from Fig.~\ref{fig:limaye} and the corresponding streamfunction 
$\psi^\infty$ from $-\int v^{\infty}_x \ dy$.  
At $22.5^\circ$S, which is the center of the domain, 
$\beta=4.6\times10^{-12}$~m$^{-1}$~s$^{-1}$ and 
$f_0=-1.4\times10^{-4}$~s$^{-1}$.  
Thus if $L_r$ were known, Eq.~\ref{eqn:qInf} shows that specifying 
$q^{\infty}$ is equivalent to specifying $gh_b$.  

\subsection{The interaction field}
\label{sec:decomposition}
Let $q(x,y)$ be a steady solution of 
QG Eqs.~\ref{eqn:qg}--\ref{eqn:q} that consists of an 
anomalous patch of potential vorticity embedded in a 
far-field flow that is zonal and steady.  
We decompose $q$ into three components:
\begin{equation}
q(x,y) \equiv q^{\infty}(y)+q^{GRS}(x,y)+q^{INT}(x,y).  
\label{eqn:qComponents}
\end{equation}
The superposition of $q^{GRS}$ and $q^{\infty}$ is not an exact 
solution because the far-field flow is deflected around the patch.  
We define the interaction potential vorticity $q^{INT}$ to represent the 
deflection of flow such that the total $q$ given by 
Eq.~\ref{eqn:qComponents} is an exact solution of the QG equations.  
Note that by definition, $q^{INT}$ asymptotes to zero both far from and 
near the patch.  
We define the interaction streamfunction and velocity to be:
\begin{equation} 
q^{INT}(x,y) \equiv \left(\nabla^2-1/L_r^2\right)\psi^{INT}(x,y)
\label{eqn:interaction}
\end{equation}
\begin{equation} 
\vec{v}^{INT} \equiv {\bf\hat{z}}\times{\bf\nabla}\psi^{INT}.
\label{eqn:vinteraction}
\end{equation}
With these definitions, the total velocity \vec{v} and its streamfunction 
$\psi$ are superpositions of the near, interaction, and far-field
components:
\begin{equation}
\psi = \psi^{\infty}+\psi^{GRS}+\psi^{INT}
\label{eqn:psivel}
\end{equation}
\begin{equation}
\vec{v} = \vec{v}^{\infty}+\vec{v}^{GRS}+\vec{v}^{INT}.
\label{eqn:vel}
\end{equation}
Note that in the linear relationships between the potential vorticity and 
streamfunction given by Eqs.~\ref{eqn:qInf},~\ref{eqn:justgrs}, 
and~\ref{eqn:interaction}, it is only the far-field component that 
contains the inhomogeneous bottom topography and $\beta$ terms.  

\section{Model definition}

\subsection{Model for far-field $q^\infty$}
\label{sec:modelFarfield}

Laboratory experiments \citep{sommeria89, solomon93} and numerical 
simulations \citep{cho96, marcus00} show that if the weather layer is 
stirred and allowed to come to equilibrium, the potential vorticity organizes 
itself into a system of east-west bands.  
The bands have approximately uniform $q$ and are separated by steep 
meridional gradients of $q$.  
The meridional gradients are all positive (i.e., have the sign as $\beta$) 
so that $q(y)$ monotonically increases from the south to the north pole 
like a ``staircase''.  
The corresponding $\vec{v}^\infty$ has alternating eastward-going and 
westward-going jet streams, with eastward-going jet streams occurring at 
every gradient or ``jump''.  
Recent measurements \citep{read06a} of the Jovian $q^\infty$ are not 
entirely consistent with this picture, for they show gradients near 
{\it both} eastward-going and westward-going jet streams.  
We therefore model $q^\infty$ between $30^\circ$S and $15^\circ$S by:
\begin{equation}
q^{\infty}(y) \equiv \sum_{i=1}^2 
\frac{\Delta Q_i}{2} \left(\tanh{\frac{y-y_i}{\delta_i}} + 1\right).
\label{eqn:qInfModel}
\end{equation}
The jumps for this model occur at $y_i$ and have strength $\Delta Q_i$, 
where $\Delta Q_i$ can be positive or negative.  
The strictly positive $\delta_i$ are a measure of the steepness of 
each jump.  
For all results in this paper, the jump locations were fixed at 
$y_1=26.0^\circ$S and $y_2=20.0^\circ$S, which are near the jet streams 
in Fig.~\ref{fig:limaye}.  
The free parameters for the model are $\Delta Q_i$ and $\delta_i$ 
for i=1,2. (Models with up-to four jumps near each jet stream were 
also tested, and the results were consistent with the ones presented here.) 

\subsection{Model for near-field $q^{GRS}$}
\label{sec:modelVortex}

We model the spatially compact $q^{GRS}$ as a piecewise-constant function 
obtained by the superposition of $M$ nested patches of uniform potential 
vorticity.  The patches are labelled $i=1,2,\cdots,M$ from innermost to 
outermost patch.  
The principal E--W diameter of a patch is denoted by $(D_x)_i$, and 
we define $q^{GRS}_i$ such that $q^{GRS}=q^{GRS}_1$ within the boundary 
of the innermost patch ($i=1$), $q^{GRS}=q^{GRS}_2$ between the boundary of 
the innermost patch ($i=1$) and the boundary of the next larger 
patch ($i=2$), and so on.  
The free parameters for the model are $M$, $q^{GRS}_i$, and $(D_x)_i$ 
for $=1,2,\cdots,M$. 
Once the free parameters for $q^{GRS}$ and $q^\infty$ are specified, 
along with the value of $L_r$, the iterative method given in appendix~B can 
be used to compute the interaction-field such that the total $q$ 
is a steady solution of the governing equations.  
Note that the {\it shapes} of patch boundaries are not free but are also 
computed by the iterative method.  

\section{Determination of best-fit parameter values}
\label{sec:bestfit}

\subsection{Decoupling of N--S velocity traits from far-field $q^\infty$}
\label{sec:decoupling}

Here we show that the N--S velocity traits are insensitive to 
the far-field potential vorticity described by Eq.~\ref{eqn:qInfModel}.  
Fig.~\ref{fig:decoupling} shows a model computed using the 
iterative method in appendix~B for the parameter values 
given in Table~\ref{table:bestFitParams}.  
The middle column of Fig.~\ref{fig:decoupling} shows that for the 
E--W velocity along the N--S axis, all three velocity components, 
$\vec{v}^{\infty}$, $\vec{v}^{GRS}$, and $\vec{v}^{INT}$, contribute 
significantly.  
However, the rightmost column shows that for the N--S velocity along the 
E--W axis, $\vec{v}^{\infty}$ has no contribution (by definition), 
and the contribution of $\vec{v}^{INT}$ is negligibly small\footnote[2]{  
The contribution of $\vec{v}^{INT}$ is small because $q^{INT}$ comprises 
two highly (E--W)--elongated slivers north and south of the GRS 
(first column, bottom row of Fig.~\ref{fig:decoupling}).  
The associated $\vec{v}^{INT}$ follows highly (E--W)--elongated 
closed streamlines approximately concentric to $q^{INT}$.  
Therefore, along the E--W axis, $\vec{v}^{INT}$ is primarily in the 
E--W direction.}.  
Only $\vec{v}^{GRS}$ contributes significantly.  
We therefore conclude that the N--S traits depend primarily on parameters 
associated with $q^{GRS}$ and are insensitive to parameters associated 
with $q^\infty$.  
This decoupling leads to a logical order for determining the best-fit 
parameter values.  
The ordering is given in Table~\ref{table:ordering} and begins with 
the determination of $(D_x)_1$ from the N--S traits.  
A more rigorous justification for the ordering is given in 
appendix~C. 

\begin{figure}
\includegraphics[width=16.5cm]{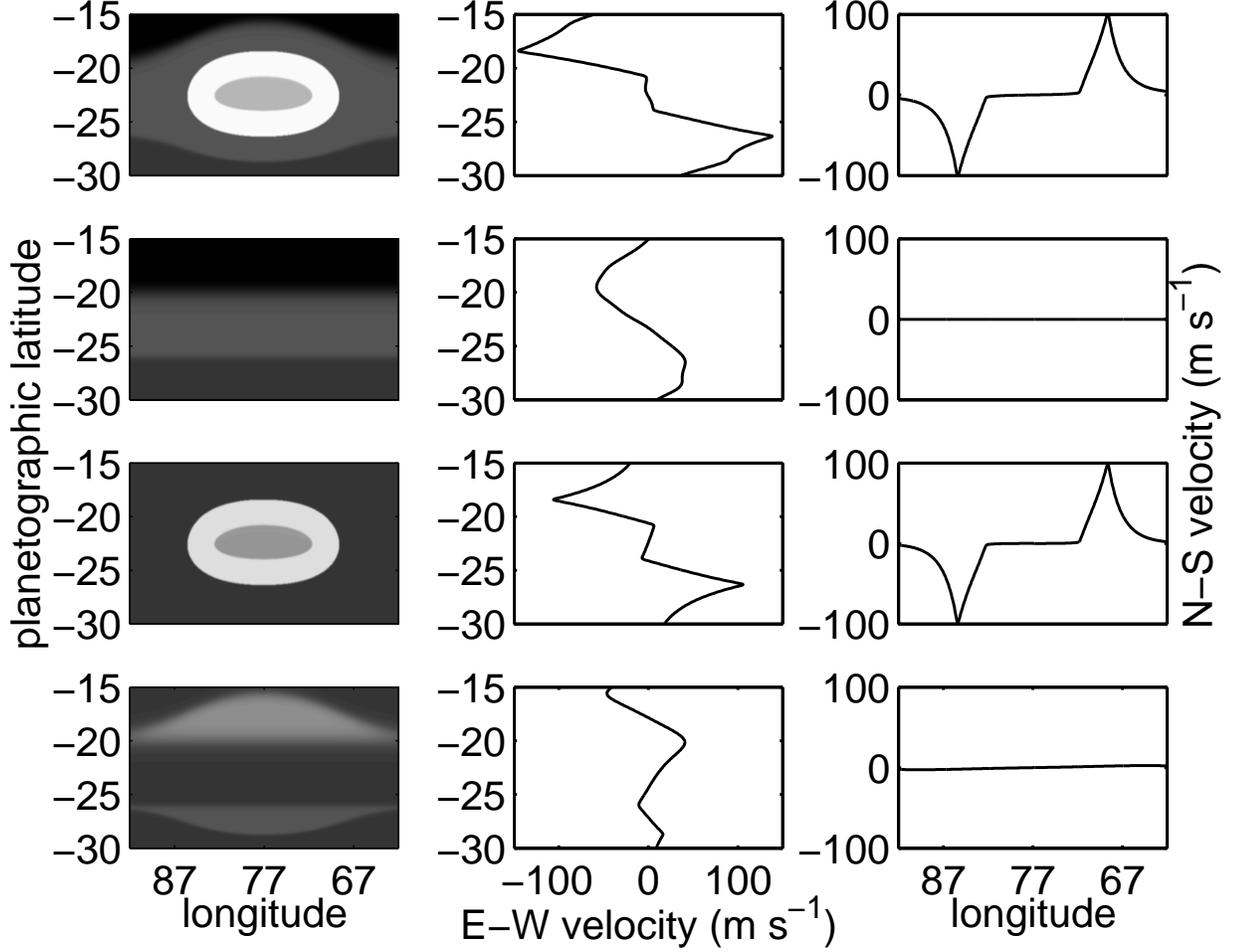}
\caption{(left to right) Potential vorticity, 
E--W velocity along the N--S axis, and N--S velocity along the E--W axis.  
A gray-scale map is used for potential vorticity with black representing 
the most cyclonic fluid and white the most anticyclonic fluid.  
(top to bottom) The components due to the total $q$, 
the components due to $q^{\infty}$ alone, the components due to $q^{GRS}$ 
alone, and the components due to $q^{INT}$ alone.  
The total $q$ is a (uniformly translating) solution of the QG Equations.  
It was computed for the parameter values in Table~\ref{table:bestFitParams} 
using the iterative method in appendix~B.  
Note that the slivers of $q^{INT}$ (left column bottom row) are due 
to the displacement of the ``jumps'' or steep gradients in $q^{\infty}$ 
as they follow streamlines that deflect around the GRS.
\label{fig:decoupling}}
\end{figure}

\begin{table}
\caption{Relationship between an observable trait of the Great Red Spot 
and the model parameter that it determines.  
The ordering of the table, from top to bottom, is the order in which 
each model parameter is determined.  
A rigorous justification for the ordering is given in appendix~C. 
\label{table:ordering}}
\begin{center}
\begin{tabular}{|l|l|}
\hline
Observable trait & Model Parameter\\
\hline
Distance between peaks in N--S velocity along E--W axis& 
E--W diameter of GRS's potential vorticity: $(D_x)_1$\\
Magnitude of peak N--S velocity along E--W axis& 
Family of possible $q^{GRS}_1$ and $L_r$\\
N--S velocity along E--W axis for $|x|\geq(D_x)_1/2$&
Unique $q^{GRS}_1$ and $L_r$ from family\\
N--S velocity along E--W axis for $|x|\leq(D_x)_1/2$& 
GRS's interior potential vorticity: $q^{GRS}_2$, $(D_x)_2$\\
E--W velocity along N--S axis & 
Far-field potential vorticity: $\Delta Q_i$, $\delta_i$\\
\hline
\end{tabular}
\end{center}
\end{table}

\subsection{Determination of best-fit $L_r$ and $q^{GRS}$ from 
N--S velocity traits}
\label{sec:NSTraits}

Here we show that an $M=2$ model is sufficient to capture the N--S 
velocity traits to within the observational uncertainties.  
For brevity, the terms {\it interior} and {\it exterior} are used in 
reference to the regions $|x|<(D_x)_1/2$ and $|x|>(D_x)_1/2$ respectively.

\begin{figure}
\begin{center}
\includegraphics[width=16.5cm]{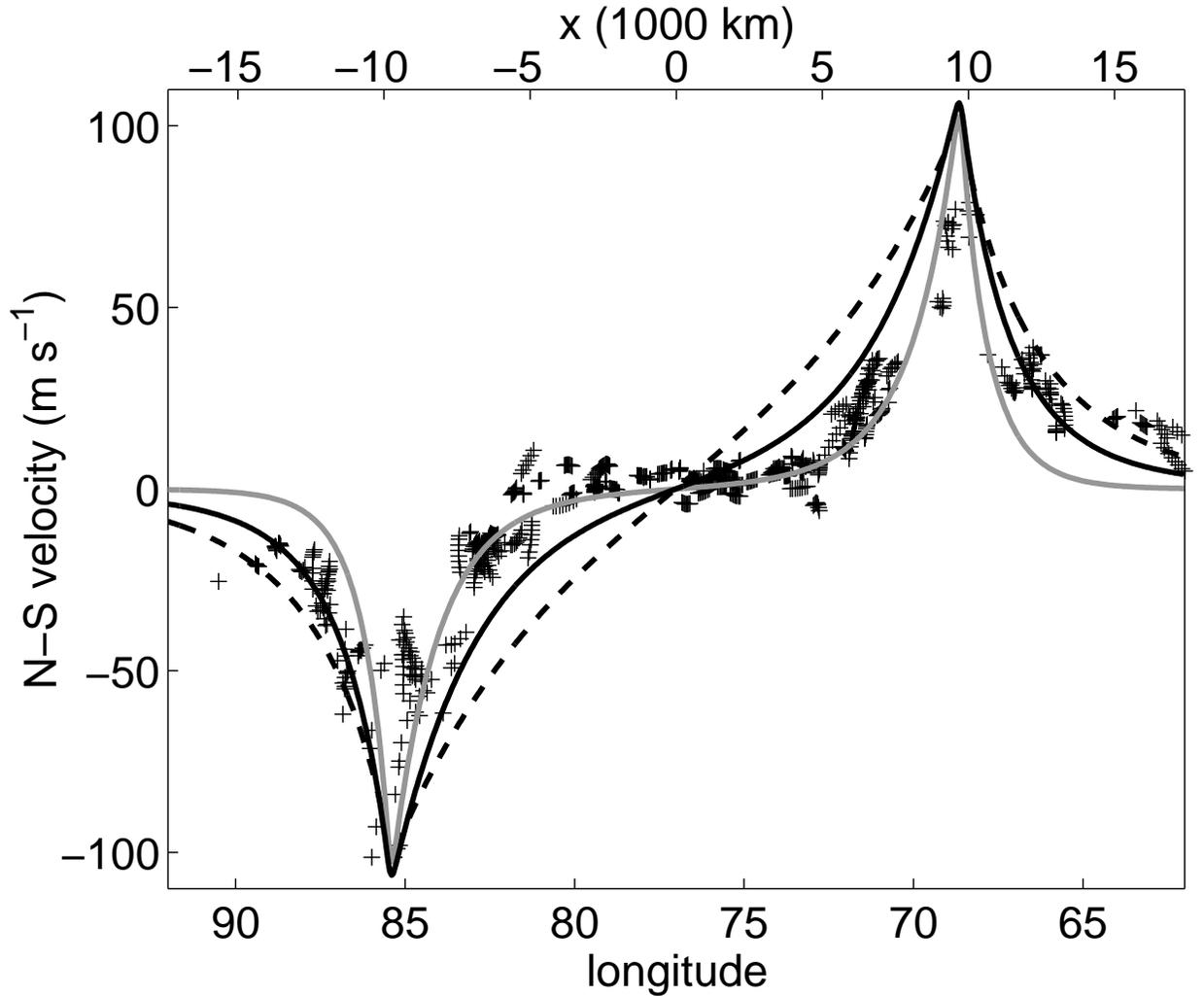}
\caption{
N--S velocity along the E--W axis for models with $M=1$.   
Crosses are {\it Voyager}~1 data from Fig.~\ref{fig:axesTiePoints}a.  
Solid black curve has $q^{GRS}_1=10.5\times10^{-5}$~s$^{-1}$ 
and $L_r=2400$~km.  
Dashed curve has $q^{GRS}_1=6.5\times10^{-5}$~s$^{-1}$ and $L_r=3800$~km.  
Solid gray curve has $q^{GRS}_1=19.5\times10^{-5}$~s$^{-1}$ and $L_r=1300$~km.  
All models have the best-fit value of $(D_x)_1=19500$~km and were 
computed using the iterative method in appendix~B.  
The N--S velocity along the E--W axis is insensitive to parameters of the 
far-field because of decoupling as described in $\S$5a.  
\label{fig:NSVelocity_Lr}}
\end{center}
\end{figure}

For $M=1$, $q^{GRS}$ is a patch of uniform potential vorticity.  
Models were computed for different values of $(D_x)_1$, $q^{GRS}_1$, 
and $L_r$.  For each model, the peaks of the N--S velocity 
along the E--W axis were found to occur at $x=\pm(D_x)_1/2$.  
The best-fit value of $(D_x)_1=19500$~km was therefore inferred from 
trait~1 in $\S$2c.  
Next, the best-fit values of $q^{GRS}_1$ and $L_r$ were constrained 
using the observed peak magnitude $V^{NS}_{\rm max}$ of the N--S 
velocity along the E--W axis.  
In particular, for a given value of $L_r$, the value of $q^{GRS}_1$ 
was chosen so that the model reproduced the observed peak magnitude.  
By repeating this process for several values of $L_r$, a 
two--parameter family (i.e., a function of $L_r$ and $q^{GRS}_1$) of 
models that simultaneously capture the observed peak locations and 
peak magnitude was obtained.  
Some family members are shown in Fig~\ref{fig:NSVelocity_Lr}.  
Note that the models do not capture the observed width of 
the northward and southward-going jets.  
In particular, for sufficiently small $L_r$ ($L_r=1300$~km), the 
model captures the rate of velocity fall-off in the interior 
but not in the exterior.  
For sufficiently large $L_r$ ($L_r=2400$~km) the opposite is true.  
For other values of $L_r$, the rate of fall-off is too fast or too 
slow in both regions.  To overcome this shortcoming, models with $M=2$ 
were considered.  

\begin{figure}
\begin{center}
\includegraphics[width=16.5cm]{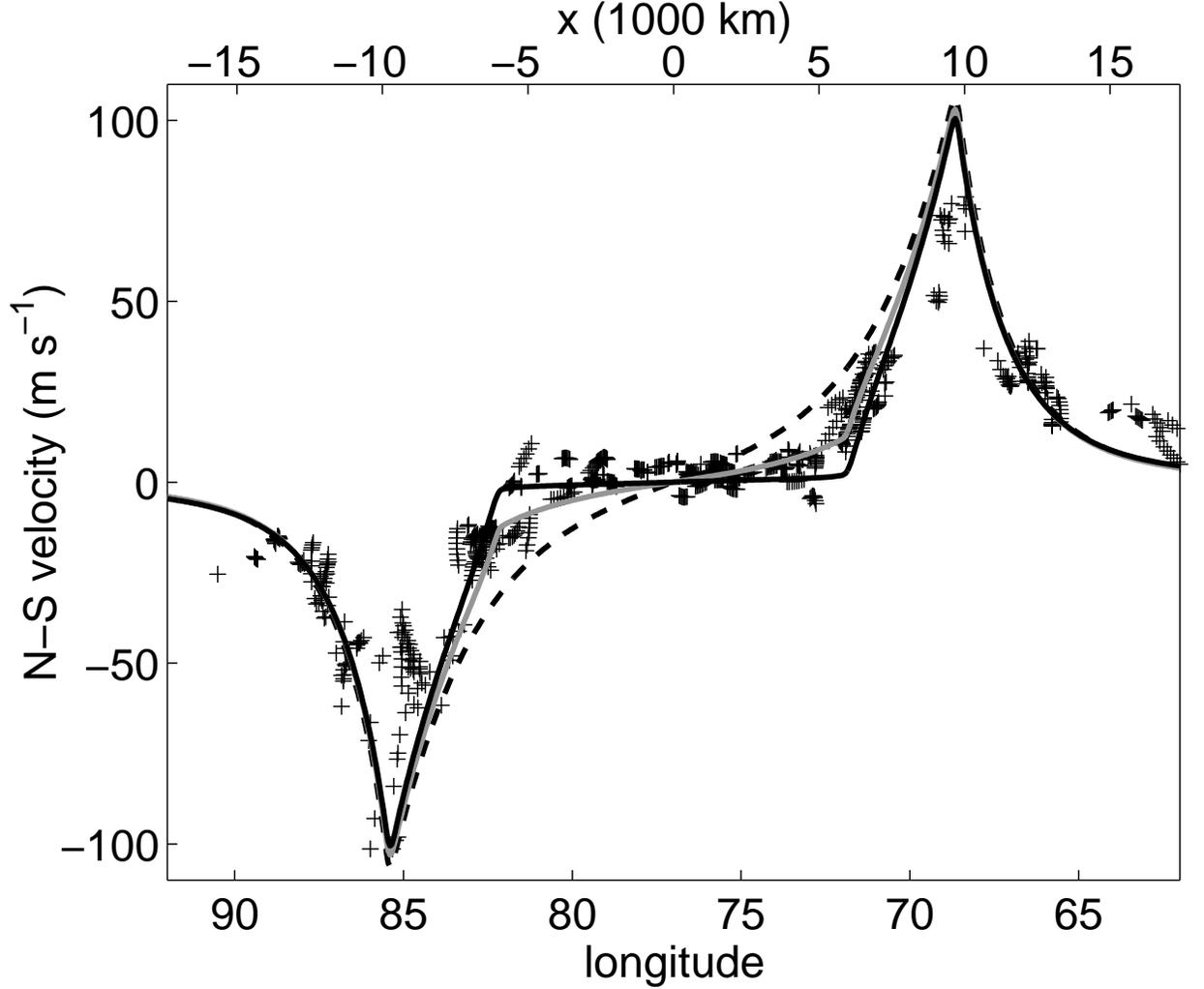}
\caption{
N--S velocity along the E--W axis for models with $M=2$.  
Each model has $L_r$, $(D_x)_1$, $q^{GRS}_1$, and $(D_x)_2$ 
set to the best-fit value in Table~\ref{table:bestFitParams}, but the 
values of $q^{GRS}_2$ for each model are different.  
The dashed curve has $q^{GRS}_2=q^{GRS}_1$, the solid gray curve has 
$q^{GRS}_2=0.8q^{GRS}_1$ and the solid black curve has the best-fit value 
of $q^{GRS}_2=0.57q^{GRS}_1$.  
Crosses are {\it Voyager}~1 data from Fig.~\ref{fig:axesTiePoints}a.  
The N--S velocity along the E--W axis is insensitive to parameters of the 
far-field because of decoupling as described in $\S$5a.  
\label{fig:NSVelocity_hf}}
\end{center}
\end{figure}

\begin{figure}
\begin{center}
\includegraphics[width=16.5cm]{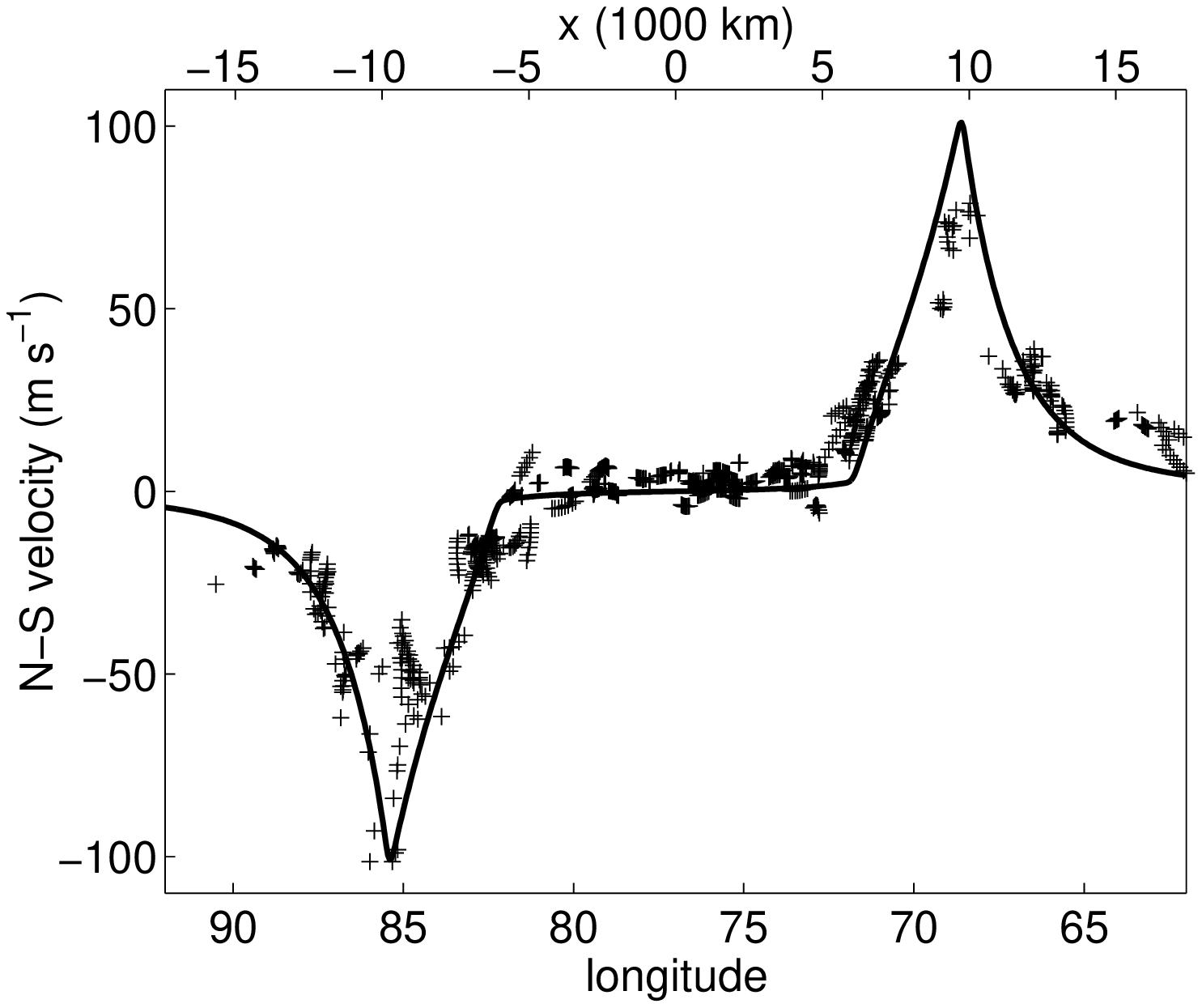}
\caption{Solid line is the N--S velocity along the E--W axis for a 
model with best-fit parameter values given in Table~\ref{table:bestFitParams}.  
Crosses are {\it Voyager}~1 data from Fig.~\ref{fig:axesTiePoints}a.
\label{fig:NSVelocity_bestFit}}
\end{center}
\end{figure}

For $M=2$, $q^{GRS}$ is the superposition of two nested patches of uniform 
potential vorticity.  
Best-fit values of $L_r$, $q^{GRS}_1$, and $(D_x)_1$ were taken from 
the $M=1$ model in Fig.~\ref{fig:NSVelocity_Lr} that captures 
the velocity profile in the exterior.  
As shown in Fig.~\ref{fig:NSVelocity_hf}, for $q^{GRS}_2=q^{GRS}_1$, 
the $M=2$ model does not capture the velocity profile in the interior.  
However, if $q^{GRS}_2$ is changed holding all other parameters fixed, 
only the interior flow changes (provided $(D_x)_1-(D_x)_2 \geq 2L_r$).  
Thus with all other parameters fixed, a genetic algorithm 
\citep{zohdi03} was used to determine values for $q^{GRS}_2$ and 
$(D_x)_2$ that minimize the $L_2$-norm difference between the model 
N--S velocity along the E--W axis and the observed velocity in the interior.  
The parameter values obtained are listed in Table~\ref{table:bestFitParams} 
and Fig.~\ref{fig:NSVelocity_bestFit} shows that the N--S traits are 
captured for these parameter values.  (Models with $M=3,4$ were also tested, 
and the results were consistent with the ones presented here.) 

\begin{table}
\caption{Best-fit parameter values for $L_r$, $q^{GRS}$, and $q^\infty$.
\label{table:bestFitParams}} 
\begin{center}
\begin{tabular}{|c|c|}
\hline
Parameter & Best-fit value \\
\hline
$L_r$ & 2400 km \\
$M$ & 2 \\
$q^{GRS}_1$ & $10.5 \times 10^{-5}$~s$^{-1}$ \\
$q^{GRS}_2$ & $ 6.0 \times 10^{-5}$~s$^{-1}$ \\ 
$(D_x)_1$ & 19500 km \\
$(D_x)_2$ & 12000 km \\
$\Delta Q_1$  & $ 1.9 \times 10^{-5}$~s$^{-1}$ \\
$\Delta Q_2$  & $-5.6 \times 10^{-5}$~s$^{-1}$ \\
$\delta_1$ &  300 km \\
$\delta_2$ & 1000 km \\
\hline
\end{tabular}
\end{center}
\end{table}

\subsection{Determination of best-fit $q^\infty$ from E--W velocity traits}

\begin{figure}
\begin{center}
\includegraphics[width=16.5cm]{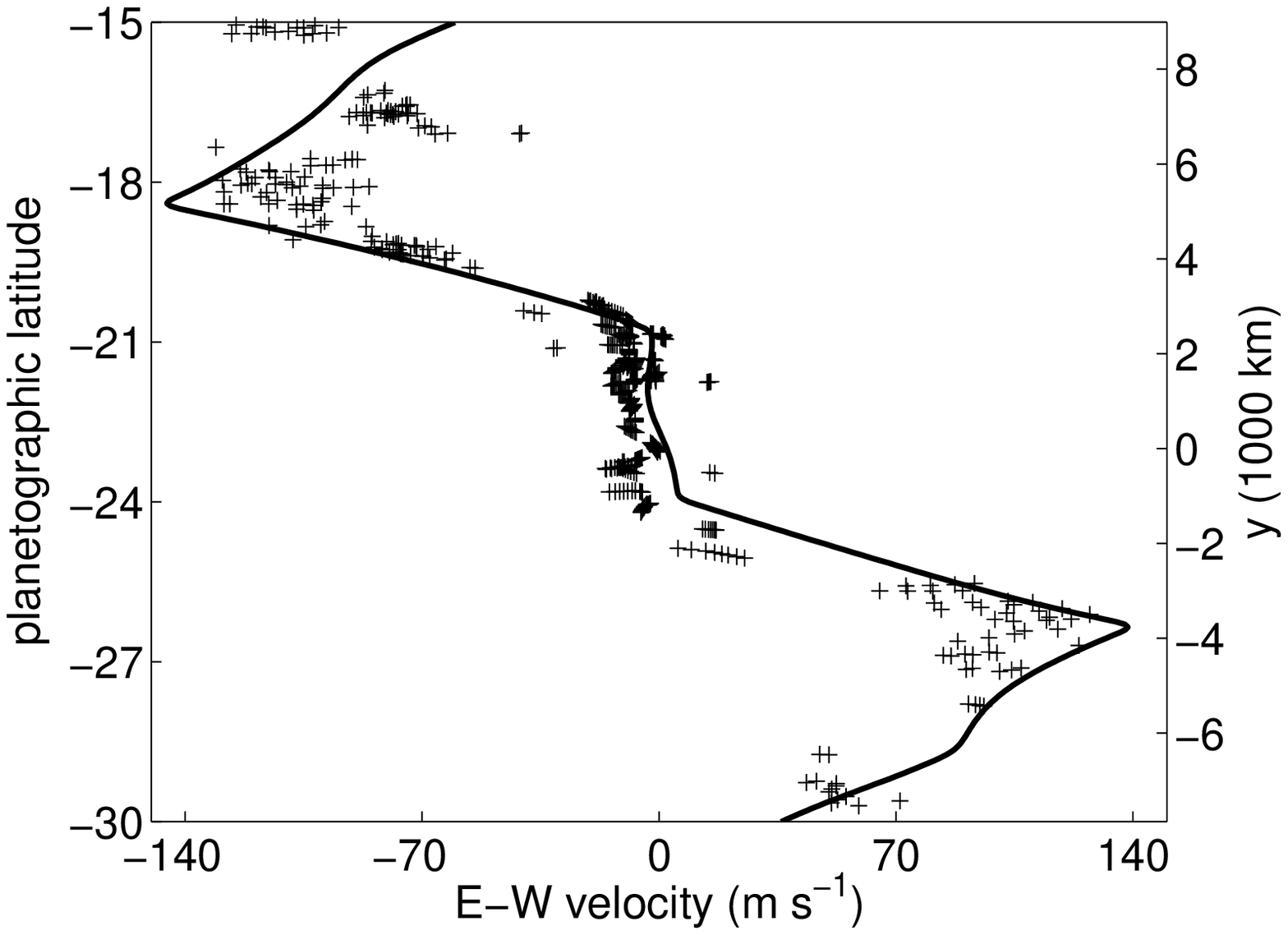}
\caption{Solid line is the E--W velocity along the N--S axis for a 
model with best-fit parameter values given in Table~\ref{table:bestFitParams}.  
Crosses are {\it Voyager}~1 data from Fig.~\ref{fig:axesTiePoints}a.
\label{fig:EW_bestFit}}
\end{center}
\end{figure}

\begin{figure}
\begin{center}
\includegraphics[width=16.5cm]{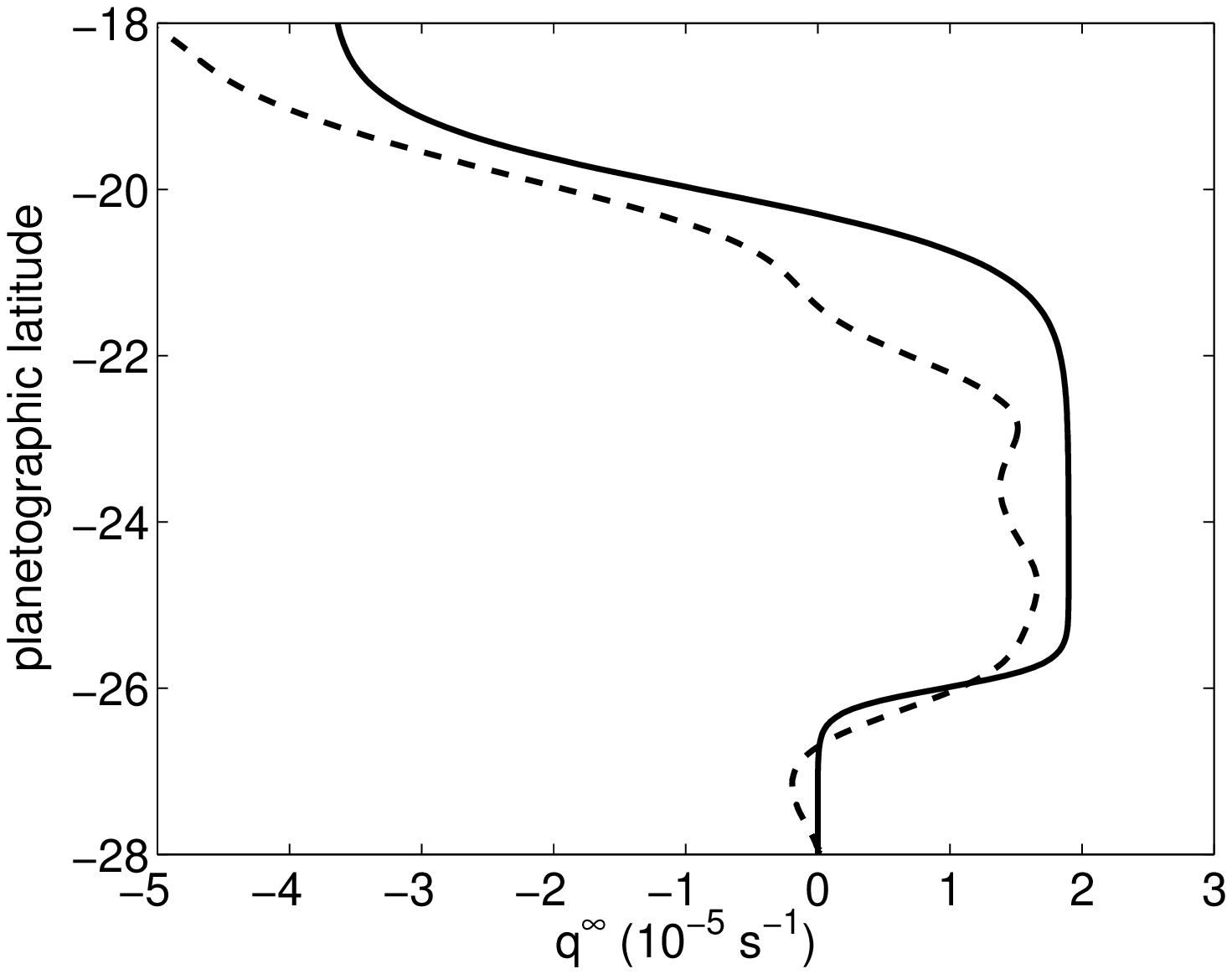}
\caption{
Solid line is far-field potential vorticity $q^\infty$ for 
model with best-fit parameters given in Table~\ref{table:bestFitParams}.  
Dashed line is $q^\infty$ determined in \citet{dowling89}.  
\label{fig:qInfinity}}
\end{center}
\end{figure}

The best-fit $q^\infty$ was determined from the E--W velocity traits.  
In particular, with the $q^{GRS}$ parameters fixed at their best-fit 
values from the preceding section, a genetic algorithm \citep{zohdi03} 
was used to determine values for $\Delta Q_i$ and $\delta_i$ that 
minimize the $L_2$-norm difference between the model E--W velocity 
along the N--S axis and the corresponding observed velocity.  
The parameter values obtained are listed in Table~\ref{table:bestFitParams}.  
Fig.~\ref{fig:EW_bestFit} shows that the E--W traits are captured for 
these parameter values.  
The corresponding $q^{\infty}$ is shown in Fig.~\ref{fig:qInfinity}.  
The velocities for this trait-capturing model were found to match 
the GRS velocities in Fig.~1 to within the observational uncertainties.

\section{Physical implications of best-fit model}

\subsection{Cloud morphology and GRS's potential vorticity anomaly}

The models show that the peak north-south velocities along the 
principal east-west axis occur at $x=\pm(D_x)_1/2$, where $(D_x)_1$ is 
the principal east-west diameter of the GRS's potential vorticity 
anomaly.  Thus the best-fit value of $(D_x)_1=19500$~km was inferred 
from trait 1 in $\S$2c.  In fact, the models show that not just the 
east-west extremeties, but the entire boundary of the GRS's potential 
vorticity anomaly is demarcated by the locations of peak velocity 
magnitude ($|{\bf v}|$).  
This implies that an estimate of the area and aspect ratio of the 
GRS's potential vorticity anomaly can be made directly from the observed 
velocity field without first determining a best-fit model.  
Traditionally, the clouds associated with the GRS have been used to infer 
the area and aspect ratio of the vortex. The east-west diameter of the 
cloud cover 
associated with the GRS is $\approx$24000--26000~km in length, which is
$\approx$25\% longer than the east-west extent of the potential vorticity 
anomaly as determined by our best-fit model.  
The north-south diameter of the cloud cover is also $\approx$25\% longer 
than that of the anomaly, so any estimate of the size of the GRS based 
on its cloud images rather than on its velocity overestimates the 
area of the potential vorticity anomaly by $\approx$50\%.  

\subsection{Rossby deformation radius}

The models show the rate of fall-off of the north-south velocity 
in the outer portion of the high-speed collar to be almost independent 
of all parameters with the exception of $L_r$.  The models also show 
the magnitude of peak north-south velocity $V^{NS}_{\rm max}$ along 
the principal east-west axis to be approximately equal to the product of 
$L_r$, the potential vorticity $q^{GRS}_1$ in the collar, and a 
dimensionless number that depends weakly on $(D_x)_1$/$L_r$ (see Table C1).  
Since $(D_x)_1$ is known from the separation of the north-south peaks, and 
$V^{NS}_{\rm max}$ can be measured to within $\pm7$~m~s$^{-1}$, the 
best-fit values of $L_r=2400$~km and $q^{GRS}_1=10.5\times10^{-5}$~s$^{-1}$ 
were determined simultaneously by demanding that the model reproduce 
the value of $V^{NS}_{\rm max}$ as well as the velocity fall-off in 
the outer portion of the collar.  

\subsection{Hollowness of GRS's potential vorticity anomaly}

The models show a uniform potential vorticity anomaly to be inconsistent 
with the north-south velocity along the east-west axis in the GRS's 
high-speed collar.  
In particular, an anomaly with uniform potential vorticity cannot 
simultaneously capture the different rates at which the velocity falls-off 
in the inner and outer portion of the collar.  
However, a model with core potential vorticity $q^{GRS}_2\approx$60\% of 
the collar's potential vorticity $q^{GRS}_1$, is able to capture both fall-off 
rates to within the uncertainties.  The ``hollowness''\footnote[3] 
{We define a hollow vortex to be one in which the absolute value of 
potential vorticity $|q|$ has a minimum in the vortex core.  
Note that hollowness is not defined in terms of vorticity $\omega$; 
vortices with uniform $|q|$ have an $|\omega|$ minimum in their cores.} 
of the GRS's potential vorticity anomaly is surprising because other 
Jovian vortices such as the White Ovals, as well as other 
geophysical vortices such as the Antarctic Stratospheric Polar Vortex 
and Gulf Stream Rings, have uniform $|q|$ or a $|q|$ maximum in their cores.  
Furthermore, hollow vortices are sometimes unstable \citep{marcus90}.  
Initial-value simulations show that hollow vortices re-distribute their $q$ 
on an advective time scale so that the final $|q|$ is uniform or has a 
maximum in the core.  
This raises the question of how a hollow GRS is stabilized over 
time scales longer than an advective time scale.

\subsection{Non-staircase far-field potential vorticity}

The best-fit $q^\infty$ has a positive jump $\Delta Q_1$ of magnitude 
$1.9\times10^{-5}$~s$^{-1}$ near the eastward-going jet stream and a larger, 
negative jump $\Delta Q_2$ of magnitude $5.6\times10^{-5}$~s$^{-1}$ 
near the westward-going jet stream.  
Due to these opposing jumps, $q^\infty$ in this region does not 
monotonically increase with $y$.  This is surprising because numerical 
and laboratory experiments (see $\S$4a) predict a monotonically 
increasing ``staircase'' profile, with jumps only at eastward-going 
jet streams.  
The best-fit profile determined here agrees qualitatively with 
a profile determined in \citet{dowling89} using an independent method.  

\subsection{Aspect ratio of GRS's potential vorticity anomaly}

The aspect ratio of the GRS's potential vorticity anomaly is 
defined to be $(D_x)_1/(D_y)_1$, where $(D_x)_1$ and $(D_y)_1$ are the 
principal north-south and east-west diameters of the anomaly respectively.  
(Recall that the shape of the the GRS's $q$ anomaly, and $(D_y)_1$ 
in particular, are obtained as output from the iterative method in appendix~B.)
The aspect ratio of the anomaly depends on the ratio of $q^{GRS}_1$ to 
the shear of the ambient flow in which the GRS is embedded; a larger 
$q^{GRS}_1$ to ambient shear ratio implies a rounder vortex while a 
smaller ratio implies a more elongated vortex \citep{marcus90}.   
It should be emphasized that, in general, the ambient shear at the location 
of the GRS is {\it not} identical to the shear of the 
far-field flow $\vec{v}^{\infty}$.  
Instead, as shown in Fig.~\ref{fig:decoupling}, the ambient shear is 
determined by the {\it interaction} of the GRS with the far-field flow.  
In particular, the middle column of Fig.~\ref{fig:decoupling} shows 
that $\vec{v}^{INT}$ is large and produces a shear with half the 
magnitude and {\it opposite} sign to the shear of $\vec{v}^{\infty}$.  
Therefore, the effect of $\vec{v}^{INT}$ is to greatly reduce the ambient 
shear at the location of the GRS.  
For the best-fit model, the aspect ratio of the anomaly is 2.18.  
If the mitigating effect of $\vec{v}^{INT}$ on the shear is eliminated 
by setting $\Delta Q_i=0$, with all other parameters, in particular 
$(D_x)_1$, unchanged from their best-fit values, then the GRS's anomaly 
shrinks in the north-south direction (i.e., $(D_y)_1$ decreases) so that 
its aspect ratio is increased by $\approx$28\%.  

The panel in the left column and bottom row of Fig.~\ref{fig:decoupling} 
explains the functional dependence of $\vec{v}^{INT}$ on $y$ and why 
its shear is {\it adverse} to the local shear of $\vec{v}^{\infty}$.  
The panel shows that the effect of deflecting the jet streams and 
associated isocontours of $q^\infty$ around the GRS is equivalent to 
placing nearly semi-circular patches of $q$ north and south of the GRS.  
When the isocontours of $q^{\infty}$ that are deflected south of the 
GRS have latitudinal gradient $d q^{\infty}/d y > 0$, the semi-circular 
patch of $q$ south of the GRS produces anticyclonic shear 
at the latitude of the GRS.  Similarly, if the isocontours that 
are deflected north of the GRS have $d q^{\infty}/d y > 0$, 
the semi-circular patch of $q$ north of the GRS produces 
cyclonic shear at the latitude of the GRS.  
Thus if the eastward-going and westward-going jet streams of $\vec{v}^\infty$, 
which are deflected respectively south and north of the GRS, both 
had $d q^{\infty}/d y > 0$, then the two semi-circular patches of 
vorticity in Fig.~\ref{fig:decoupling} would have opposite sign and 
form a dipole.  
The dipole would create a large net westward flow at the latitude 
of the GRS, but would create little shear (none, if the patches 
had equal strength) there.  However, for the best-fit model,  
the westward-going jet stream has $d q^{\infty}/d y < 0$ and the 
eastward-going jet stream has $d q^{\infty}/d y > 0$.  
Both semi-circular patches are anticyclonic and the result is a large 
shear that is adverse to the shear of $\vec{v}^{\infty}$, as shown 
in Fig.~\ref{fig:decoupling}.  

\section{Conclusions and Future Work}

In this paper we have described a technique for determining quantities of 
dynamical significance from the observed velocity fields 
of a long-lived Jovian vortex and its neighboring jet streams.  
Our approach was to model the flow using the simplest governing equation and
the fewest unknown parameters that would reproduce the observed 
velocity to within its observational uncertainties.
For the {\it Voyager}~1 data, this is a nine-parameter model that is an 
exact steady solution to the 1.5--layer reduced gravity QG equations.  
The nine parameters are the local Rossby deformation radius, 
the $q$ in the GRS's high-speed collar, the $q$ in the GRS's core, 
the east-west diameter of the GRS's $q$ anomaly, 
the east-west diameter of the GRS's core, the size and steepness of 
two jumps in the far-field $q$, one located near the latitude of the 
eastward-going jet stream to the south of the GRS, and the other located 
near the westward-going jet-stream to its north.  
We determined ``best-fit'' values for the nine parameters by 
identifying several ``traits'' of the observed GRS velocity field 
and seeking a model that reproduced all those traits.  

Perhaps the most surprising result of our study was that the simple model 
described above was able to reproduce the entire observed velocity field 
in Fig.~1 to within the uncertainties of 7\% (that is, 7~m~s$^{-1}$).  
The success of the model is due, in part, to the fact that the GRS must 
be well-described by the QG equations, and to the fact that 
the model is an exact steady solution of the governing equations.  
The success is also due to the fact that the chosen traits are robust 
and in some sense unique (e.g., hollowness) to the physics associated 
with the GRS.  Finally, a part of the success of the model is due to the 
relatively large uncertainties (7\%) of the {\it Voyager}~1 velocities 
compared to more recent data sets (see below). 

Our most important result was to show that the interaction between the 
GRS and its neighboring jet streams determines the shape of the GRS's $q$ 
anomaly.  
By explicitly computing the interaction, we showed that the effect of the 
GRS is to bend the jet steams (identified by their jumps in $q$) 
so that they pass around the GRS, and the effect of the bending of the 
jet streams is to reduce the zonal shear at the location of the GRS.  
A smaller zonal shear at the location of the GRS compared to the $q$ of the 
GRS implies a smaller east-west to north-south aspect ratio for the GRS's 
$q$ anomaly.  
The best-fit model has a positive jump at the eastward-going jet stream 
and a larger, negative jump at the westward-going jet stream.  
The bending of these opposing jumps significantly reduces the 
zonal shear at the GRS, making the aspect ratio of the GRS's $q$ 
anomaly $\approx$28\% smaller (i.e., rounder) than it would be if there were 
no interaction with the jet streams.  
It is also interesting to note that due to the opposing jumps, 
the far-field $q$ does not monotonically increase from south to north, 
which is contrary to numerical and laboratory experiments that predict 
a monotonically increasing ``staircase'' profile.  

The GRS's potential vorticity anomaly was found to be ``hollow'' with core 
potential vorticity $\approx$60\% that of the collar; this is curious 
because hollow vortices are generally unstable.  
The locations of peak velocity magnitude were found to be accurate markers 
of the boundary of the GRS's $q$ anomaly, which implies that the area and 
aspect ratio of the anomaly can be inferred directly from the velocity data.  
On the other hand, clouds associated with the GRS are not an accurate 
marker of the anomaly as they differ from the anomaly area by $\approx$50\%. 
This suggests that cloud aspect ratios, areas, and morphologies should not 
be used to determine temporal variability of Jovian vortices.

In devising the model, our philosophy was to include no more complexity 
than was required to match the observed velocity to within its uncertainties.  
However, lower-uncertainty measurements of the velocity field using 
CIV \citep{asaydavis06} on observations from {\it Hubble Space Telescope}, 
{\it Cassini}, and {\it Galileo}, may require that the QG approximation 
be relaxed in favor of shallow-water.  
Also, if thermal observations \citep{read06b} are to be accounted 
for, governing equations that permit 3D baroclinic effects will be required.  
Modeling different data sets would show how the best-fit parameter values 
evolve with time. 

A companion paper to this one shows that the best-fit model is stable and 
explores the stabilizing effects of the hollow GRS--jet stream interaction.
Demonstrating stability is important because hollow vortices 
are usually unstable.  
Finally, there are several questions raised by our best-fit model of the GRS 
that will need to be answered. How did a hollow GRS form?
Why are there no other hollow Jovian vortices (for which the velocity has
been measured, cf. the current Red Oval and the three White Ovals 
at 33$^{\circ}$S, which existed between the mid-1930's and 1998)? 
One possible answer to the second question is that Jovian vortices 
apart from the GRS lack opposing jumps near their neighboring jet streams 
and the associated reduction in shear due to the vortex--jet stream 
interaction.  Indeed, a preliminary best-fit model of the 
White Ovals \citep{shetty06} does not show opposing jumps near the 
neighboring jet streams.  

\begin{acknowledgements}
We thank the NASA Planetary Atmospheres Program for support.  
Computations were run at the San Diego Supercomputer Center 
(supported by NSF).  
One of us (PSM) also thanks the Miller Institute for Basic Research in 
Science for support.  
\end{acknowledgements}

\newpage
\appendix

\section{Method for removing spurious velocities and correcting 
for the curvature of cloud trajectories}

The method involves two stages of iteration.  
We start with the velocity from \citet{mitchell81} in which the 
trajectories are assumed to be straight lines and the velocities are 
assumed to be located mid-way between tie-point pairs (the initial and 
final coordinates of a cloud feature in a pair of images is 
defined to be a tie-point pair).  
We then spline the irregularly spaced 
tie-point velocities onto a uniform grid of size $0.05^\circ\times0.05^\circ$.  
The first step of the inner loop of the iteration computes, for each 
tie-point pair, 
the {\it curved} trajectory that a passive Lagrangian particle would follow 
beginning at the initial
tie-point location ($x_I, y_I$) to its final tie-point location ($x_F, y_F$), 
using a $5^{th}$-order Runge-Kutta integration.
To carry out the integration, the velocity field is spline-interpolated from 
the grid to the off-grid locations where it is required by the 
integrating scheme. The integration creates a set of closely spaced points,
$(x_i,y_i)$, $i=1,2,3,\cdots,N$, along the trajectory, where 
$(x_1,y_1) \equiv (x_I, y_I)$.  In general, this trajectory does not end 
with $(x_N,y_N)$ equal to $(x_F,y_F)$ as desired.  
We therefore compute a second trajectory $(X_i,Y_i)$ starting from the 
final tie--point location $(x_F, y_F)$ and integrate backward in time using 
the interpolated velocity from grid points. 
A third trajectory 
$(\bar{x}_i,\bar{y}_i)\equiv[(N-i)(x_i,y_i)+(i-1)(X_i,Y_i)]/(N-1)$ is
then computed as a linear interpolation that, by construction, starts at 
$(x_I,y_I)$ and ends at $(x_F,y_F)$.  Moreover, because
the points along each trajectory are close together, the velocity at 
each point  $(\bar{x}_i,\bar{y}_i)$
is well-approximated with the temporal, second-order finite difference 
derivative using the nearest neighbor trajectory points.
A new velocity at the grid points is created from the
spline of the velocities along the curved trajectories of all of the 
tie-point pairs (for each trajectory, we use the velocities at eight 
approximately equally spaced points along the trajectory).  
We then return to the first step of the inner loop.  
We use the original set of tie-point pairs, but the velocity is now the 
updated velocity on the grid. The inner loop is iterated until it 
converges (typically, three iterations).
We then compute the residual of each velocity vector, which is defined to be 
the magnitude of the difference between the original, uncorrected 
tie-point velocity and the converged velocity interpolated by splines 
to that location.  Velocity vectors with residuals that were six times the 
root-mean-squared value of all of the residuals were considered to 
be spurious, and their tie-points were removed from the original data set.  
Once the spurious points are removed, the outer loop is complete and the 
entire process is repeated starting with the new (diminished) set of 
tie-points.  The outer loop was iterated until no more tie points were 
removed.  
The {\it Voyager}~1 tie-point set required three iterations of the outer 
loop and resulted in the removal of 220 of the original 1100 points.  
The root-mean-squared residual of the iterated velocity is 
$\approx7$~m~s$^{-1}$, and we use this value as a measure of the 
uncertainty in the data. 
For comparison, it should be noted that the residual of the {\it Voyager}~1 
tie-points without correcting for curvature and without removing spurious
tie-points is $\approx9$~m~s$^{-1}$, and the residual for the Hubble 
Space Telescope data (from CIV) for the GRS using the method described here is 
$\approx3$~m~s$^{-1}$ \citep{asaydavis06}.  
In the high-speed collar, the residuals in the vorticity derived by 
differentiating the {\it Voyager}~1 velocity are $\approx35\%$ of 
the maximum vorticity.  

\section{Iterative method for computing steady-state solutions of 
the 1.5--layer reduced gravity QG equations}
\label{sec:algorithm}

Here we describe an iterative method for computing steady solutions of 
Eqs.~\ref{eqn:qg}--\ref{eqn:q} subject to periodic boundary 
conditions\footnote[4]{
While periodicity is natural in the east-west direction, it is 
artificially imposed in the north-south direction.  This is done by 
embedding the domain of interest 
(where the velocities are designed to match those of Jupiter) 
into one with 20\% larger latitudinal extent.  
The flow velocities in the northern and southern extremities 
of the enlarged domain do not match those of Jupiter, but smoothly interpolate 
the velocities from the domain of interest to the periodic boundaries.  
} in $x$ and $y$.
The method seeks solutions that consist of a single anomalous patch 
embedded in a zonal flow, and that are steady when viewed in a frame 
translating with the patch.  
Such solutions are of the form $q(x,y,t)=q(x-Ut,y)$, where $U\bf{\hat{x}}$ 
is the constant drift velocity of the vortex.  
Substituting for $q$ in Eq.~\ref{eqn:qg} we obtain:
\begin{equation}
({\vec{v}}-U{\bf\hat{x}})\cdot{\vec{\nabla}} q=0,
\label{eqn:isocontour}
\end{equation}
which implies that isocontours of $q$ and isocontours of 
$\psi+U y$ are coincident.  
It is this property that the iterative method exploits to compute 
uniformly translating solutions.
As input, the method requires $L_r$, $q^{GRS}_i$, $(D_x)_i$ for 
$i=1,2,\cdots,M$, and $\Delta Q_i$, $\delta_i$ for $i=1,2$.  
As output, the method provides $q^{INT}$, $U$, and the shape of each 
vortex patch.  
Initial guesses must be supplied for the quantities obtained as output.  
The guesses are then iterated, keeping the input quantities fixed, 
until the total $q$ is a uniformly translating solution.  
The iterative procedure is described below.  
The domain is $x\in[-L_x,L_x]$, $y\in[-L_y,L_y]$.  
The origin is at the point of intersection of the principal axes.  

\begin{itemize}

\item[1.]A new $\psi$ is computed from the current $q$ 
by inverting the Helmholtz operator in Eq.~\ref{eqn:q}.  
The current $q$ is the sum of $q^\infty$, the current $q^{GRS}$, 
and the current $q^{INT}$.  

\item[2.]A new drift speed $U$ for the anomaly is computed.  
The drift speed of the anomaly, as derived in \citet{marcus93}, 
is given by $\int_A q^{GRS}v_x dA/\int_A q^{GRS} dA$, 
where $A$ is the current area of the anomaly.  

\item[3.]New isocontours of $(\psi + U y)$ are computed.  The 
isocontours are streamlines of the current velocity in the 
translating frame.  Streamlines that extend from the western to the 
eastern boundary of the domain are referred to as {\it open} streamlines.  
Streamlines that are not open are referred to as {\it closed}.  

\item[4.]A new $q^{INT}$ is computed.  This is done by 
setting the value of $q$ along an open streamline to the value 
of $q^\infty$ at the point on the western boundary through 
which the streamline passes.  
In other words, if $y=s(x)$ is the equation of an open streamline, then 
$q(x,s(x))\equiv q^\infty(s(-L_x))$ for $x \in [-L_x,L_x]$. 

\item[5.]A new $q^{GRS}$ is computed by computing a new boundary for patches 
$i=1,2,\cdots,M$.  The new boundary is identified as the closed 
streamline that passes through $x=(D_x)_i/2$.  
Note that if the current patch is reflection symmetric about the 
N--S axis, the value of $(D_x)_i$ is conserved.  
The potential vorticity of each patch $q^{GRS}_i$ is held fixed.  

\end{itemize} 

The iterations are repeated until $U$ converges to within a desired tolerance, 
or equivalently, until isocontours of $(\psi+Uy)$ and isocontours of $q$ are 
coincident.  For all calculations in this paper, the initial guess 
for the shape of a patch was an ellipse with $(D_y)_i=0.5(D_x)_i$.   
The final shapes are reflection symmetric about the 
N--S axis, but they are not symmetric about the E--W axis.  
The initial $q^{INT}$ and $U$ were set to zero. 
The grid resolution was $0.05^\circ\times0.05^\circ$.  
The equilibria are not sensitive to the domain size provided the 
domain boundaries are at least three deformation radii away from the edge 
of the outermost patch.  
We note that it would be interesting to explore initial guesses that 
are not reflection symmetric about the N--S axis, to see if asymmetry 
persists for the final solution.  
Indeed, recent low-uncertainty measurements of the GRS velocity field 
\citep{asaydavis06} show asymmetry about the N--S axis.  
For the {\it Voyager}~1 data set however, any asymmetry is 
much smaller than the uncertainties, so asymmetric models are deferred to 
future work.  

\section{Sensitivity of model traits to model parameters}
\label{sec:uncertainty}

Here we quantify the sensitivity of a model trait to small changes 
in a model parameter.  
The results justify the methodology used in $\S$5 to determine the 
best-fit parameter values.  
Consider a trait of the N--S velocity along the E--W axis, 
say the peak magnitude $V^{NS}_{\rm max}$.  
From dimensional analysis it is rigorous to write:
\begin{equation}
V^{NS}_{\rm max} = L_r q^{GRS}_1 \, F[q^{GRS}_2/q^{GRS}_1, 
\Delta Q_1/q^{GRS}_1, \Delta Q_2/q^{GRS}_1, \delta_1/L_r, \delta_2/L_r, 
(D_x)_1/L_r, (D_x)_2/L_r], 
\label{eqn:da}
\end{equation}
where $F$ is a dimensionless function of seven dimensionless arguments 
(note that Eq.~\ref{eqn:da} is completely general if the value of 
$V^{NS}_{\rm max}$ is independent of $\vec{v}^{\infty}$ as is 
suggested by decoupling; otherwise, and in particular, for any trait of the 
E--W velocity along the N--S axis, the function $F$ would have to 
include arguments of the dimensionless scalars that parametrize 
$\vec{v}^{\infty}$).  
The sensitivity of $V^{NS}_{\rm max}$ to changes in a particular parameter, 
say $L_r$, was determined by computing the value of $V^{NS}_{\rm max}$ 
for a change in $L_r$ of $\pm$5\% around its best-fit value with all other 
parameters fixed at their best-fit values, and then using a finite difference 
scheme to construct the dimensionless partial derivative 
$(L_r/ V^{NS}_{\rm max}) \,  (\partial V^{NS}_{\rm max} /\partial L_r)$ 
$\equiv$ $\partial \, ln \, V^{NS}_{\rm max} / \partial \, ln \, L_r$.  
Dimensionless partial derivatives computed for other traits 
are listed in Table~\ref{table:sensitivity}.  
We consider a trait to be {\it insensitive} to any parameter for which the 
absolute value of its dimensionless partial derivative is much less than 
unity.  Note that the results are consistent with Table~1. 

The partial derivatives are not independent. For example, four 
of the parameters in Eq.~\ref{eqn:da} have dimensions of inverse time 
(and we write them as $\tau_i$, $i=1, 2, 3, 4$), and 
four have dimensions of length (and we write them as $\chi_i$, $i=1, 2, 3, 4$).
Differentiation of Eq.~\ref{eqn:da} yields the following constraints:
\begin{equation}
\sum_{i=1}^{4} \partial \, ln  \, V^{NS}_{\rm max} /\partial ln \, 
\tau_i \equiv 1, \label{eqn:partialt}
\end{equation}
and
\begin{equation}
\sum_{i=1}^4 \partial \, ln \, V^{NS}_{\rm max} /\partial ln \, \chi_i \equiv 1. \label{eqn:partiall}
\end{equation}
In general, a trait $L$ that has dimensions of length, such as the width of 
the N--S jet, and which depends only on the parameters in Eq.~\ref{eqn:da}, 
must satisfy the following constraints:
\begin{equation}
\sum_{i=1}^{4} \partial \, ln \, L /\partial \, ln \, \tau_i \equiv 0, 
\label{eqn:dLdT}
\end{equation}
and
\begin{equation}
\sum_{i=1}^4 \partial \, ln \, L /\partial \, ln \, \chi_i \equiv 1. 
\label{eqn:dLdL}
\end{equation}
Table~\ref{table:sensitivity} shows that all traits with the exception of 
$(D_y)_1$ satisfy the constraints.  
The reason $(D_y)_1$ does not satisfy the constraints is that it is a 
trait of the E--W velocity and therefore also depends on parameters 
associated with the far-field flow $\vec{v}^\infty$.  

The uncertainties in the best-fit parameter values may be quantified as 
follows.  The $L_2$ norm difference between the best-fit 
velocity and the velocity in Fig.~1 is computed.  The $L_2$ norm difference 
is then recomputed with all parameters fixed at their best-fit values 
with the exception of parameter $L_r$ (say).  
A curve of the $L_2$ norm difference as a function of $L_r$ is 
then computed.  By construction, the curve has a minimum at the best-fit 
value of $L_r$.  The width of the curve at half-minimum is identified as 
the uncertainty in $L_r$.  
Since measurements of the GRS velocity using CIV have much lower 
uncertainties than the {\it Voyager} velocity and will soon be 
available \citep{asaydavis06}, we did not deem it useful to compute 
parameter uncertainties for the analyses in this paper.  

\begin{table}
\caption{Sensitivity of model traits to model parameters.  
Each entry is 
(parameter/trait) $\, \times \,$ $\partial$(trait)/$\partial$(parameter) 
$ \equiv$  $\partial \, ln \, $(trait)/$\partial \, ln \, $(parameter).  
The partial derivatives were evaluated at the best-fit parameter values 
in Table~\ref{table:bestFitParams}.  
The full-width-at-half-maximum ($W$) for the northward-going jet 
along the principal east-west axis is broken into two pieces 
$W \equiv E+I$.  $E$ (or $I$) is the distance along the east-west 
axis between the location where the north-south velocity reaches its 
maximum value of $V^{NS}_{\rm max}$, and the location in the outer 
(or inner) portion of the high-speed collar where the north-south 
velocity falls to $V^{NS}_{\rm max}/2$.  
Since the models are reflection symmetric about the principal 
north-south axis (due to the reflection symmetric initial guess, 
see appendix~B), the northward-going and southward-going jets have 
identical values of $V^{NS}_{\rm max}$, $I$, and $E$.  
\label{table:sensitivity}}
\begin{center}
\begin{tabular}{|l|c|c|c|c|c|c|c|c|c|}
\hline
Model trait $\downarrow$ / Model parameter $\rightarrow$ 
& $(D_x)_1$ & $L_r$ & $q^{GRS}_1$ & $q^{GRS}_2$ & $(D_x)_2$ & 
$\Delta Q_1$ & $\Delta Q_2$ & $\delta_1$ & $\delta_2$ \\ 
\hline
Peak N--S velocity along E--W axis, $V^{NS}_{\rm max}$ & 
0.3 & 1.0 & 1.1 & 0.0 & -0.2 & 0.1 & -0.1 & 0.0 & 0.0\\
Distance between N--S peaks along E--W axis & 
1.0 & 0.0 & 0.0 & 0.0 & 0.0 & 0.0 & 0.0 & 0.0 & 0.0\\
Exterior width of N--S jets at half maximum, $E$ & 
0.1 & 1.0 & 0.1 & 0.0 & -0.1 & 0.0 & 0.0 & 0.0 & 0.0\\
Interior width of N--S jets at half maximum, $I$ & 
1.1 & 0.5 & 0.0 & -0.1 & -0.7 & 0.1 & -0.1 & 0.0 & 0.1\\
N--S diameter of GRS's potential vorticity, $(D_y)_1$ & 
0.6 & 1.1 & 0.4 & -0.1 & -0.3 & 0.1 & -0.3 & 0.0 & 0.0\\
\hline
\end{tabular}
\end{center}
\end{table}

\bibliographystyle{ametsoc} 
\bibliography{references}

\end{document}